\documentclass[12pt]{article}
\usepackage[dvips]{graphicx}
\usepackage{amssymb,amsfonts,amsmath,bm}
\usepackage[top=1.05in,bottom=1.05in,left=1in,right=1in]{geometry}

\begin{document}
\title{Universal driving structure of self-sustained oscillatory
complex networks}
\author
{Xuhong Liao,$^{1\ast}$ Weiming Ye,$^{1\ast}$ Xiaodong Huang,$^{1}$ Qinzhi Xia,$^{1}$ \\
 Xuhui Huang,$^{1}$ Pengfei Li,$^{1}$ Yu Qian,$^{1}$ Xiaoqing Huang,$^{1}$ Gang
 Hu$^{1\dag}$\\
\\
\normalsize{$^{1}$Department of Physics, Beijing Normal University,}\\
\normalsize{Beijing 100875, China,}\\
\normalsize{$^\ast$ These authors contributed equally to this work.}\\
 \normalsize{$^\dag$To whom correspondence should be
addressed; E-mail: ganghu@bnu.edu.cn} }
\date{}
\maketitle

{\bf Recently, self-sustained oscillations in complex networks
consisting of nonoscillatory nodes (network oscillators) have
attracted great interest in diverse natural and social fields. Due
to complexity of network behaviors, little is known so far about the
basic structures and fundamental rules underlying the oscillations,
not to mention the principles of how to control it. In this article
we propose a common design principle for oscillations; predict novel
and universal Branched Circle (BC) structures of oscillatory
networks based on this principle; and suggest an operable Complexity
Reduction Method to reveal the BC structures. These ideas are
applied to excitable cell networks (including neural cell networks),
and genomic regulatory networks. Universal BC structures are
identified clearly in these two considerably different systems.
These BC structures reveal for the first time both oscillation
sources and wave propagation pathways of complex networks, and guide
us to control the oscillations with surprisingly high efficiency.
\bf}

\section*{Introduction}
Self-sustained oscillations in complex networks consisting of
non-oscillatory nodes are very popular phenomena in natural and
social systems. These oscillations are extremely important in
controlling various basic rhythms in wide fields, such as
oscillatory neural networks (1--7), sinoatrial node rhythms in
cardiac systems (8--10), oscillatory cycles in genomic regulations
(11--17) and so on. Though the topic of self-sustained oscillations
has been investigated for some decades, many fundamental questions
remain in puzzle. For instance,we do not know whether there are some
common principles underlying the oscillatory behaviors of complex
networks in diverse fields; whether there are some common structures
hidden in the complicated interaction schemes that determine the
dynamics of oscillations; and if these common structures do exist,
how one can find them. Specially, an given oscillatory networks as
Figs.\ 1{\it A}, 1{\it B}, one can hardly say anything about where
the oscillation sources are; how oscillatory waves propagate from
the sources to the whole networks and how one can efficiently
control the oscillations based on these understandings. None of
these questions of crucial importance has been answered if networks
are sufficiently complicated.

In this article we have made the following essential advances. (i)
We start from a simple while solid basis design principle: any
individually nonoscillatory node can oscillate if and only if it is
driven by one or few interactions with advanced phases. (ii) Based
on this design principle, all periodically oscillatory
one-dimensional (1D) networks must have universal unidirectional
coupling structures of branched circles (BCs, Fig.\ 2) in the form
of circles with radiating branches. (iii) Oscillatory
high-dimensional complex networks can be reduced to 1D BC networks
by applying the method of dominant phase-advanced driving pathes.

We further apply the above ideas to both oscillatory excitable cell
networks (ECNs, neural cell networks included) (5--7,18--20) and
genomic regulatory networks(GRNs) (21--24). It's the first time the
oscillations of these two considerably different systems are studied
with a unified approach. We have reached the following results. (i)
We apply the same design principle and dimension reduction approach
to two kinds of systems and obtain the same universal BC structures;
(ii) From the BC patterns of both systems we can clearly identify
the oscillation sources (unidirectional regulatory circles) and
reveal the phase propagation pathways (unidirectional tree
branches); (iii) With these BC patterns we can classify one or few
most important nodes, by controlling which we can control the
oscillations of the whole networks with surprisingly high
efficiency.

\section*{Design principles and universal structures}

Figure 1{\it A} considers an ECN example in which multiple cells
with a large number of random couplings show a rather complicated
interaction pattern. Although each individual node is
nonoscillatory, with certain initial conditions we observe periodic
oscillations, one of which is shown in Fig.\ 1{\it B}. We further
study how sensitive the oscillation is to the control. After
exhaustive tests we find that the oscillation can be terminated
(i.e., turned to the homogeneous rest state $u_i=v_i=0, i=1,2,..,N$)
by removing a single red node (node $78$) shown in Fig.\ 1{\it C},
while the oscillation persists safely if we remove as many as $70$
empty square nodes in Fig.\ 1{\it C}. Here removing a node we mean
to discard all interactions from and toward the given node. The
results of Fig.\ 1{\it C} are highly surprising. They clearly show
that though the topological interaction structure of Fig.\ 1{\it A}
is random and homogeneous, the dynamic organization supporting
self-sustained oscillation is strongly heterogeneous, and the roles
played by different nodes in this organization are considerably
different. The central task of the present work is to reveal these
self-organized patterns under the conditions of full knowledge of
coupling structure and output data and then to achieve an effective
control of the oscillatory networks based on this understanding.

Considering a network with N nodes, dynamic variables are associated
to each node, and these variables obey well defined coupled ordinary
differential equations (ODEs). Each node is nonoscillatory
individually while the entire complex networks are periodically
oscillatory. Regardless of different dynamics and coupling forms for
different systems, we propose a common design principle for such
oscillatory networks.

Design principle: each nonoscillatory node can oscillate if and only
if it is driven by one or few oscillatory interactions with advanced
phases.

The definitions of ``advanced phase" for different systems will be
explained later. Let us first consider the simplest 1D oscillatory
network with each node phase-advancedly driven by one node only
(networks with N nodes and N unidirectional interactions). Suppose
an arbitrary node $i_1$ is phase-advancedly driven by a node $i_2$
via coupling, which is phase-advancedly driven by node $i_3$ in
turn, and this successive unidirectional driving chain goes as
$i_1\leftarrow i_2\leftarrow i_3\leftarrow \cdots \leftarrow
i_k\leftarrow \cdots$. Since N is finite we must come to a node
$i_q$, $q\leq N$, which is driven by one of the previous nodes
$i_1,i_2,...,i_{q-1}$, say $i_p$ ($p<q$). Then a successive
regulatory loop $i_p\leftarrow i_{p+1}\leftarrow \cdots \leftarrow
i_q\leftarrow i_p$ is formed, serving as the oscillation source of
all other nodes. Therefore, all 1D oscillatory networks must have
the structure of circles with radiating branches, schematically
shown in Fig.\ 2.

The pattern of Fig.\ 2 gives a picture of a branched circle (BC),
and it is thus called as BC structure. This structure is universal
for self-sustained periodic oscillations in 1D networks consisting
of nonoscillatory nodes. Since no nonoscillatory node oscillates
without phase-advanced driving from other nodes, two key rules must
be obeyed by any BC structure:

  (i)  There must be few (at least one) successively phase-advanced
  driving circles.

  (ii) Each node not in the circles must be in a tree branch rooted
  at a node in a circle.

The BC circle is obviously the oscillation source without which the
network can never oscillate, and the tree branches show various
pathways of phase propagations starting from different circle nodes.
All nodes in Fig.\ 2 can be classified according to their locations
in the BC pattern. We expect that the circle nodes controlling large
branches may be of the most importance for the oscillation. In
comparison all nodes near the branch ends with few or even without
downstream nodes have the lowest influences on the oscillation of
the network. We will show later that these expectations are well
confirmed by numerical results with large probability.

 The simple and instructive scheme of Fig.\ 2 is deduced in 1D
phase-advanced driving networks. However, interaction structures of
complex networks in general (e.g., Fig.\ 1{\it A} which are
high-dimensional and random) are much more complex than Fig.\ 2.
Therefore, we propose an operable and physically meaningful method
to reduce original random patterns (as Fig.\ 1{\it A}) to the simple
and instructive BC pattern of Fig.\ 2. The method consists of the
following Complexity Reduction (CR) steps:

(a) Find phase-advanced driving interactions for each node.

(b) Find the single dominant interaction among these phase-advanced
    driving interactions.

(c) Use all these dominant interactions to unidirectionally link the
network nodes, and draw the dominant phase-advanced path pattern,
which turns out to be the 1D BC pattern of Fig.\ 2.

All the above steps are generally applicable in diverse fields for
self-sustained oscillations of complex networks of individually
nonoscillatory nodes. The particular meanings of ``advanced phase"
and ``dominant phase-advanced driving" should be properly defined,
according to realistic physical, chemical and biological interaction
mechanisms in each individual system.

\section*{Coupled excitable cell networks}

 We now consider complex excitable cell networks (ECNs) of the B\"{a}r Model(18)

$$\frac{du_i}{dt}=\frac{1}{\varepsilon}
u_i(1-u_i)(u_i-\frac{v_i+b}{a})+D_u \sum_{j=1}^{\nu}(u_{ij}-u_i)\
,$$
$$\frac{dv_i}{dt}=f(u_i)-v_i,\qquad i=1,2,...,N \qquad \qquad \qquad \quad \eqno (1)$$
$$ f(u_i)=\left\{
\begin{array}{cl}
0\qquad\qquad\qquad\qquad\ u_i\leq \frac{1}{3}\ ,\\
1-6.75u_i(u_i-1)^2 \quad \ \frac{1}{3}<u_i<1\ ,\\
1\qquad\qquad\qquad\qquad\ u_i>1\ .
\end{array} \right.
$$
in which $u_{ij}$ means the variable $u$ of the $j$th node linked to
node $i$. Note, two major features of neural networks are precisely
the excitability of cell dynamics and the complexity of interaction
network(20,25--27). Without couplings all cells of ECNs are not
oscillatory individually for certain given $a$, $b$, they evolve
asymptotically to the rest state $u=v=0$ and will stay there forever
unless some external force drives them from this state. Therefore,
all the analyses in the former section are applicable to this type
of systems. Whenever a cell is kicked from the rest state by a small
stimulus, the cell can oscillate by its own internal dynamics (so
called excitable dynamics). Therefore, for a given node that enters
the region of the rest state ($u<u_{th}$) at time $t_s$ and departs
from this region at time $t_e$, we define ``phase-advanced drivings"
by the interactions from those neighbors which leave from the rest
state earlier than the given node (i.e., in the period ($t_s,t_e$))
which thereby provide favorable interactions in kicking the given
cell from the rest state. Among these phase-advanced interactions
the dominant phase-advanced driving is defined by the interaction
from the node first leaving from the rest state in the period (
$t_s,t_e$). It is no doubt that the dominant driving must give the
most important contribution to excite the given node, i.e., to drive
the given node to oscillate.

For simplicity, we assume symmetric couplings, and adopt random
complex networks with identical coupling degree $\nu$ (i.e.\ each
cell couples to equal number $\nu$ of other cells). We also take
identical parameters for all cells. One advantage of this simplest
homogeneous assumption is to make sure that all self-sustained
oscillatory behaviors here are not due to any heterogeneity in
topological structure, but due to the self-organized heterogeneity
of dynamical mutual excitations.


Figure 1{\it A} shows an ECN with random mutual couplings with
identical degree $\nu=4$. With the structure of Fig.\ 1{\it A} we
simulate the system by taking different sets of random initial
conditions. In most of cases the system evolves asymptotically to
the homogeneous rest state. However, about $8\%$ of tests provide
periodic oscillations, and the state in Fig.\ 1B is one of them.

In Fig.\ 3{\it A} we plot the BC pattern draw from Figs.\ 1{\it A}
and 1{\it B} by applying the CR method. We find a BC pattern being
the type of Fig.\ 2. In Fig.\ 3{\it A} the BC circle plays the role
of oscillation source, and phase waves propagate down all the tree
branches rooted at various cells in the circle. The BC structure of
Fig.\ 3{\it A} clearly shows distinctive levels of significances of
different cells for oscillation that cannot be observed in Fig.\
1{\it A}. In Fig.\ 1{\it A} all cells stand equivalently in the
homogeneous and randomly coupled network, and no cell takes any
priority over others from the topological structure. The situation
in Fig.\ 3{\it A} is different. Cells in the circle and cells in
various turning points of large branches (which control large
numbers of downstream cells) are likely to be important to the
oscillation. In particular, node $78$ is likely to be the most
important cell because it locates in the circle on one hand and
controls large branches with a huge number of downstream nodes on
the other hand. It is interesting to observe that node $78$ is the
very single red node in Fig.\ 1{\it C} that we found important for
controlling the oscillation but did not know the reason then. In
Fig.\ 3{\it B} we show how the oscillation collapses quickly to the
homogeneous rest state after removing only a single red node $78$.
On the other hand, cells near the branch ends may be much less
significant for the oscillation. We remove simultaneously a large
number of cells ($70$ nodes) in branch tails (See Fig.\ 3{\it C})
which are exactly the empty square nodes in Fig.\ 1{\it C}, the
network not only continues its periodic oscillation (Fig.\ 3{\it
D}), but also keeps the BC structure almost unchanged for the cells
remained (Fig.\ 3{\it C}). Therefore, the question for Fig.\ 1{\it
C}, why so many empty square nodes together are much less important
than a single red cell, is clearly answered in Fig.\ 3{\it A}:
because all these empty square nodes are far from the centered
oscillation generator, and have little influence on the
self-sustained oscillation; on the other hand the single red cell
$78$ controls the oscillation source and a huge number of downstream
nodes, and it is of crucial importance for the given oscillation.
Removing one or few other cells in the circle may not stop the
oscillation but can considerably change the BC structure of the
oscillation source.

Frequency is an important quantity describing the properties of
oscillatory networks. We further study the influence of BC circles
on frequencies of networks. We computed Eqs.\ (1) by taking
different random couplings and random initial conditions for
$N=100$, $\nu=3$, and found some oscillatory realizations (all are
periodic), we then measured the frequency $\omega$ of each
oscillatory network, and plot $\omega$ vs $n$ in Fig.\ 3E with $n$
being the size of the corresponding circle. The red square
represents the average frequency $<\omega>$ and the solid line
denotes the linear fitting of the tendency. In Fig.\ 3E monotonous
and nearly linear decrease of $<\omega>$ with $n$ is clearly
demonstrated. These observations convincingly support the conclusion
that BC circles play the role of oscillation sources in complex
networks.

Figure.\ 4{\it A} is another ECN network with $N=100$ and $\nu=3$.
For certain initial condition, we observe periodic oscillation of
Fig.\ 4{\it B}. In Fig.\ 4{\it A} we show that this oscillation can
be terminated by removing a pair of nodes (red nodes $12$ and $21$)
in contrast with Fig.\ 1{\it C} where oscillation can be suppressed
by removing only a single node. Similar to Fig.\ 1{\it C}
oscillation persists when we simultaneously remove all $60$ empty
square nodes shown in Fig.\ 4{\it A}. The mystery difference between
Figs.\ 1{\it C} and 4{\it A} can be again well explained by the
corresponding BC patterns. In Fig.\ 4{\it C} we show the BC pattern
corresponding to the oscillation in Figs.\ 4{\it A} and 4{\it B}.
The essential and interesting difference between Fig.\ 3{\it A} and
Fig.\ 4{\it C} is that Fig.\ 4{\it C} contains two BC clusters
instead of the single one in Fig.\ 3{\it A}. The two red nodes shown
in the two BC circles of Fig.\ 4{\it C} are exactly identical to
those in Fig.\ 4{\it A} with the same color. Comparing the BC
patterns of Fig.\ 4{\it C} with Fig.\ 3{\it A}, we understand the
difference of Fig.\ 1{\it C} and Fig.\ 4{\it A} immediately. Since
there are two oscillation source circles in Fig.\ 4{\it C}, we have
to destroy both circle structures for terminating oscillation by
removing two key nodes in Fig.\ 4{\it A} simultaneously, each from a
circle of Fig.\ 4{\it C}, and this is sharply different from the
single circle structure of Fig.\ 3{\it A}. In Figs.\ 4{\it D} and
4{\it E} we get the BC patterns with one of the two red nodes (node
$12$ and node $21$) removed, respectively. It's interesting to see
that in both cases one BC circle of Fig.\ 4{\it C} is destroyed
while the other remained circle serves as the unique oscillation
source, and all nodes of the destroyed BC cluster are engrafted to
the survival circle for continuing their periodic oscillations. In
Fig.\ 4{\it F} we remove $60$ side cells (the empty nodes)
simultaneously from Fig.\ 4{\it C}, and find again that periodic
oscillation persists, and the corresponding BC pattern of the
unremoved nodes are not affected. We have tested a number of
different ECNs with different initial conditions, different
interaction degree and structures, different system sizes, or
different ECN models including FHN neural cell networks, and
obtained very rich behaviors of BC patterns. The general conclusions
of universal BC patterns are verified in all cases, and some of
these results are shown in Supporting Information Part $1$.

\section*{Complex genomic regulatory circuits}

We now consider another model of self-sustained oscillations of
genomic regulatory networks (GRN), the dynamics is described by the
following coupled ODEs(21--24).

$$ \frac{dx_i}{dt}=\mu_i-\gamma_i x_i+f_i, \quad
 f_i=\left\{
  \begin{array}{ll}
  A_i(\bm {x}) \quad \quad \quad &\mbox{Positive \ regulation}\ , \\
  R_i(\bm {x}) \quad \quad \quad &\mbox{Negative \ regulation}\ ,\\
  A_i(\bm {x})R_i(\bm {x}) \quad &\mbox{Joint \ regulation}\ ,
\end{array} \right.
$$
$$ \bm{x}=(x_1,x_2,\ldots ,x_N)\ , \hspace{55mm} \eqno(2)$$
$$ A_i(\bm {x})=\frac{act_i^{h_i}}{act_i^{h_i}+K_i^{h_i}}\ ,
 \quad R_i(\bm {x})=(1-\mu_i)\frac{K_i^{h_i}}{rep_i^{h_i}+K_i^{h_i}}\ ,$$

$$\quad \ act_i=\sum_{j=1}^N \alpha_j^{(i)} x_j \ ,\quad
rep_i=\sum_{j=1}^N \beta_j^{(i)}x_j \ ,\quad (i,j=1,\ldots,N) \ ,$$
where $x_i$ represents the concentration of protein corresponding to
node $i$, and $act_i$ ($rep_i$) represents the summation of
activatory (repressive) transcriptional factors. These ODEs can be
derived from a full set of  equations of both mRNA and potein
concentrations via adiabatic approximation when the time scales of
transcription and translation are separable(11). For simplicity we
consider homogeneous parameter distributions again for all nodes
($\mu_i=\mu$, $\gamma _i=\gamma$, $h_i=h$, $K_i=K$,
$\alpha_i=\beta_i=1$, $i=1,2,\ldots,N$). It's emphasized that the
GRN dynamics Eqs.\ (2) is apparently different from ECN Eqs.\ (1)
for their intrinsic node dynamics (one-variable passive dynamics for
Eqs.\ (2) against two-variable excitable dynamics for Eqs.\ (1)),
coupling dynamics (highly nonlinear positive or negative regulatory
interactions for Eqs.\ (2) while linear and diffusive coupling for
Eqs.\ (1)), and coupling structure (unidirectional couplings for
Eqs.\ (2) against symmetric couplings for Eqs.\ (1)). It is a
surprise for us to observe essentially the same dynamic BC
structures in both GRN and ECN systems as shown in the following.

Each node in Eqs.\ (2) has passive dynamics. Without coupling,
variable $x_i$ must evolve to a fixed value, and any periodic
oscillation of $x_i$ must be driven by one or few periodic
interactions from other nodes. Let us approximately simplify an
arbitrary one-variable passive dynamics with a periodical driving
$\dot{x}=\mu-\gamma x+ f(x,t)$ by linearizing the oscillatory
elements around a stable stationary solution of the autonomous
dynamics

$$ \triangle \dot{x}=-\lambda\ \triangle x+Acos(\omega t) \ , \eqno (3) $$
which has the asymptotic periodic motion

$$ \triangle x(t)=\frac{A}{\sqrt{\lambda ^2+\omega ^2}}\ cos(\omega t-\phi) \ ,$$
$$ sin\phi=\frac{\omega}{\sqrt{\lambda ^2+\omega ^2}}\ ,\quad
cos\phi=\frac{\lambda}{\sqrt{\lambda ^2+\omega ^2}}\ , \eqno(4) $$
leading to the phase-advanced driving condition of

$$ 0 \lesssim \phi \lesssim \frac{\pi}{2}\ . \eqno (5)$$
We represent the phase of node $i$ by $\phi_i$ and the phase of
interaction from node $j$ to node $i$ by $\phi_{j \rightarrow i}$,
which is identified by

$$ \phi_{j\rightarrow i}= \left\{
\begin{array}{ll}
\phi_j &\mbox{positive regulation}\ ,\\
\ \\
\phi_j+\pi &\mbox{negative regulation}\ .
\end{array}
 \right.
\eqno (6)$$ The condition of phase-advanced driving reads
$$ 0 \lesssim \phi_{j \rightarrow i}-\phi_i \lesssim \frac{\pi}{2} \ . \eqno (7)$$
With interaction structures and full periodic oscillation data of
complex GRN known we can use Eqs.\ (6) and (7) to identify all
phase-advanced interactions among all interactions (CR method (a)).
To operate the CR method (b) we simply define the dominant driving
by the interaction among all phase-advanced interactions which has
the largest oscillation amplitude. The detailed discussion on the
phase and amplitude computation is given in Supporting Information
Part $2$. After choosing the dominant interactions for various
nodes, we can link all nodes by the dominant interactions and
construct BC patterns from the original GRNs.

It is known that a necessary condition for a genomic regulatory loop
to be oscillatory is that the genes in the loop must interact
successively in a manner of negative feedback(13,21--24), i.e., the
number of negative couplings should be odd(13,22). We call a
successive unidirectional interaction loop with an odd number of
negative interactions as an oscillatory negative feedback
loop(NFL)(28) in the following. Note, existence of one or multiple
oscillatory NFLs is the necessary but not sufficient condition of
oscillatory networks. In Supporting Information Part $2$ (Fig.\ S5),
we show these oscillatory NFLs for $q \leqslant 5$. Nevertheless,
given a complex network (e.g., Fig.\ 5{\it D} or 5{\it E}), there
may exist a huge number of possible oscillatory NFLs. There is so
far no report of a method to find out which NFLs are in function to
produce a given oscillatory pattern. And this is right our following
task.


In order to examine the validity of the general prediction on
universal oscillatory BC patterns we have computed huge number of
GRNs by varying dynamic parameters (homogeneous and heterogeneous),
changing autoregulatory dynamics (with or without autoregulation,
with negative or positive autoregulation); changing the number of
nodes and the structure of couplings and also by varying the initial
variable preparations. Specifically, we have made $328400$ tests
(about $1400$ oscillatory realizations) by exhaustively computing
different coupling structures of 3-node circuits; $10^4$ random
tests for $4$-node and $5$-node circuits (totally $2 \times 10^4$
tests, $708$ oscillations); $5\times 10^3$ random tests for
$10$-node and $20$-node circuits (totally $1\times 10^4$ tests,
about $1083$ oscillations), we found that none of the tests violates
the predictions. In particular, we found that in all oscillatory
cases we can succeed in constructing BC patterns all of which have
circles being one of the oscillatory NFLs in Fig.\ S5. In Figs.\
5{\it A-E} we show some periodically oscillatory examples of
regulatory networks. Although the complicated interaction structures
don't disclose any clue of the mechanism supporting the
oscillations, by applying CR method we succeed in reducing the
original complex networks to the corresponding BC patterns greatly
simplified in Figs.\ 5{\it F-J}, respectively, which fully confirm
the prediction of Fig.\ 2. Each BC pattern in Fig.\ 5 has a source
circle being one of the oscillatory NFLs in Fig.\ S5, and all other
nodes are in various tree branches rooted at one of circle nodes,
showing wave propagation pathways.

From Figs.\ 5{\it F-J} we expect that the nodes in the circles or
near the circles may be important for the given oscillatory states
while nodes near the ends of various branches may be less
significant. We study each gene's influence on oscillations by
removing it. In Figs.\ 5{\it F-J} any node whose individual removal
results in the termination of the oscillations are filled with red
color, and empty otherwise. We find that by removing a gene on a
circle we have very large probability to terminate the oscillation.
However, when we remove a node located at the end of a branch
pathway oscillations have much larger probability to persist. For
statistics we have made a detailed investigation for $10$-node
oscillatory GRNs, and found that if we remove an arbitrary single
node on BC circles the probability to terminate oscillations is
about $84\%$ while this probability is down to about $24\%$ if an
arbitrary single node on branches is removed.

For identifying the system response to control, we study the dynamic
behavior of $N=20$ (Fig.\ 5{\it I}) in more detail. In Fig.\ 6{\it
A} we show oscillation of $<x(t)>=\frac{1}{N} \sum_{i=1}^{N} x_i(t)$
damping to a fixed value after a key circle node removed at
$t=1000$. On the other hand, the oscillation persists (Fig.\ 6{\it
B}) and the BC structure is only slightly modified (Fig.\ 6{\it C})
as a node at a branch end is removed. Most interestingly, whenever
self-sustained oscillations are maintained after removing some
nodes, the BC circles have a strong tendency to be unchanged (Fig.\
6{\it C}) or slightly modified by refinding some interaction bridges
to repair the circles (Fig.\ 6{\it D}). All these observations
verify the significance of the universal BC structures for
oscillations of complex networks.

In Eqs.\ (2) we assume ``AND" role between activators and repressors
for multiple-factor regulations(21). Some regulatory circuits may
obey ``OR" role(11,23,29). Though the coupling dynamics of ``OR"
rule looks considerably different from Eqs.\ (2), all analyses for
Eqs.\ (2) can be identically applied to the ``OR" cases. This aspect
is discussed in Supporting Information Part $3$.

\section*{Discussion}

In conclusion we study the problem of self-sustained periodic
oscillations in complex networks consisting of nonoscillatory nodes.
We propose a general design principle of oscillatory networks, based
on which we reveal that phase-advanced driving BC patterns (Fig.\ 2)
are the universal structures of simplest 1D oscillatory networks.
And complicated high-dimensional networks can be reduced to these 1D
BC patterns by applying the method of dominant phase-advanced
interactions. From the BC patterns we can easily identify
oscillation sources and phase propagation pathways of oscillatory
complex networks. All these messages are deeply hidden in the
original complex coupling structures and random phase distributions.
These BC structures are extremely important for understanding and
efficiently controlling self-sustained oscillations of complex
systems. We successfully used these ideas and methods to analyze
models of excitable cell networks and genomic regulatory circuits.
These ideas, methods and universality of structures are expected to
be applicable to self-sustained oscillations of complex networks in
broad range of fields. In recent decades, the concept and functions
of central pattern generators (CPGs) have attracted great attention
in the field of neural networks(30--33). In this article we show for
the first time how to uncover CPG-like patterns in periodically
oscillatory complex networks from complicated interaction structures
and seemingly mess phase distribution data.

In the present article we consider only cases of periodic
oscillations where all BC patterns are stationary. If oscillations
are quasiperiodic or even chaotic, BC patterns may vary during the
evolutions, and this opens a new field for the further study.
Moreover, throughout this article we study how to reveal BC patterns
with full knowledge of the interaction structures and oscillation
data. These conditions are not fulfilled in many experiments. Thus,
it is another crucial task to extend the investigations to the cases
with partial data available.

This work was supported by the National Natural Science Foundation
of China under Grant Nos. 10675020 and the National Basic Research
Program of China (973 Program)(2007CB814800).

\clearpage

\noindent {\bf Fig.\ 1.}  A complex excitable cell network of Eqs.\
(1) consisting of $N=100$ nodes. $a=0.84$, $b=0.07$,
$\varepsilon=0.04$. All these parameters will be used in Figs.\ 1,
3, 4. All couplings with strength $D_u=0.2$ are randomly chosen
between $N$ nodes each having coupling degree $\nu=4$. {\it(A)} An
example of network interaction structure. {\it(B)} Periodic orbit of
average variables $<u(t)>=\frac{1}{N}\sum_{i=1}^N u_i (t)$ and
$<v(t)>=\frac{1}{N}\sum_{i=1}^N v_i (t)$ of the network (A) for a
set of initial conditions randomly chosen. {\it(C)} Oscillation of
(B) can be terminated by removing a single red node. By removing a
node we mean to discard all interactions from and towards the given
node. However, oscillation persists if we simultaneously remove
other $70$ empty square nodes.

\noindent {\bf Fig.\ 2.} Schema of universal structure of
self-sustained oscillatory $1$D networks with all nodes
nonoscillatory individually. The arrowed lines indicate
unidirectional phase-advanced interactions. Characteristic features
are: one or few unidirectionally interacting circles serving as the
oscillation sources together with unidirectionally interacting
branches radiated from the circle showing wave propagation pathways
(called branched circles, BC).

\noindent {\bf Fig.\ 3.} BC patterns and their oscillation dynamics
of ECNs. {\it(A)} BC pattern of Eqs.\ (1) reduced from Figs.\ 1(A)
and 1(B) by applying CR method. From this pattern we are able to
identify oscillation source (the unidirectional circle) and wave
propagation pathways (the tree-like branches from various nodes of
the circle). Self-sustained oscillation of Fig.\ 1(B) can be
effectively suppressed by removing only a single red node $78$,
which is the same as that in Fig.\ 1(C). {\it(B)} Oscillation
suppression by removing node $78$ at $t=t_D=10$. {\it(C)} BC pattern
drawn after removing simultaneously the $70$ empty square nodes
shown in Fig.\ 1(C). Oscillation is kept and the BC pattern of the
remaining nodes is not affected. {\it(D)} Oscillation evolution with
$70$ empty square nodes of (C) removed at $t=t_D=10$,
simultaneously. {\it(E)} Frequency $\omega$ of a periodic complex
network of Eqs.\ (1) plotted vs the size $n$ of the BC circle. Large
red square represents the average frequency $<\omega>$, and solid
line is the linear fitting of the data. $N=100$, $\nu=3$, $D_u=1.0$.
Coupling structures and initial conditions are randomly chosen. A
strong correlation between $<\omega >$ and $n$ is demonstrated by
the monotonously decreasing curve.

\noindent {\bf Fig.\ 4.} Manipulations of oscillatory complex
networks. {\it(A)} Another ECN network with $N=100$, $\nu=3$,
$D_u=1.0$. {\it(B)} Oscillatory orbits of network (A) with a certain
set of initial condition. Oscillation persists after all the $60$
empty square nodes shown in (A) are removed simultaneously.
Apparently different from Fig.\ 1(C), now removing any single node
can no longer terminate the oscillation. Oscillation can be
suppressed by, at least, removing the pair of red nodes in (A). The
difference between Figs.\ 1(C) and 4(A) is again an interesting
mystery. {\it(C)} BC pattern of the oscillation in (B) considered.
Now we find two separated BC clusters from two circles. Removing the
pair of red nodes $12$ and $21$ can break the two source circle and
terminate the oscillation. {\it(D)} BC pattern with a single node
$12$ removed. Now one BC circle with node $12$ is destroyed and the
remained circle of (C) plays the role of the unique oscillation
source. {\it(E)} BC pattern with node $21$ removed. {\it(F)} BC
pattern after removing $60$ side nodes from (C) (i.e., the $60$
empty square nodes in (A)).

\noindent {\bf Fig.\ 5.} Complex oscillatory GRNs and the
corresponding BC patterns. The solid green (red dashed) line denotes
positive (negative) regulation. The lines in the following figures
on GRNs have the same meanings throughout the article.
{\it(A)}-{\it(E)} Some examples of $4$-node ($I=9$ interactions),
$6$-node ($I=20$), $10$-node ($I=55$), $20$-node ($I=105$),
$40$-node ($I=224$) GRNs , respectively. $\mu_i=\mu=0$,
$\gamma_i=\gamma=0.1$, $h_i=h=2$, $K_i=K=0.3$, $\alpha_i=\beta_i=1$,
$i=1,2,\ldots,N.$ With the given interaction structures all these
circuits show self-sustained periodic oscillations with arbitrary
initial conditions. {\it(F)}-{\it(J)} BC patterns constructed from
the periodic oscillation of networks of (A)-(E), respectively, by
applying CR method. In all cases we find that each BC circle is one
of the oscillatory NFLs in Fig.\ S5. A node whose individual removal
terminates the oscillation is filled with red, otherwise it is
empty.

\noindent {\bf Fig.\ 6.} Detailed numerical results of Eqs.\ (2) for
the case of Figs.\ 5(D) and 5(I) ($N=20$). {\it(A)} Time evolution
of the average concentration $<x>$, which periodically oscillates
for $t \leqslant t_D$ and collapses to a stationary solution after
 we remove a circle node $15$ at $t_D=1000$. {\it(B)(C)} Time evolution and BC
 pattern , respectively, with node $6$ removed at $t_D=1000$.
 The BC circle of Fig.\ 5(I) is not changed by removing this
 branch gene. {\it(D)} BC pattern with node $18$ removed. In (D) the
 BC circle must be changed by removing a circle gene, however,
 the system repairs the broken circle by finding some interaction
 bridges with most of original circle nodes kept on the new circle.

\newpage
{\centering\scalebox{0.8}{\includegraphics{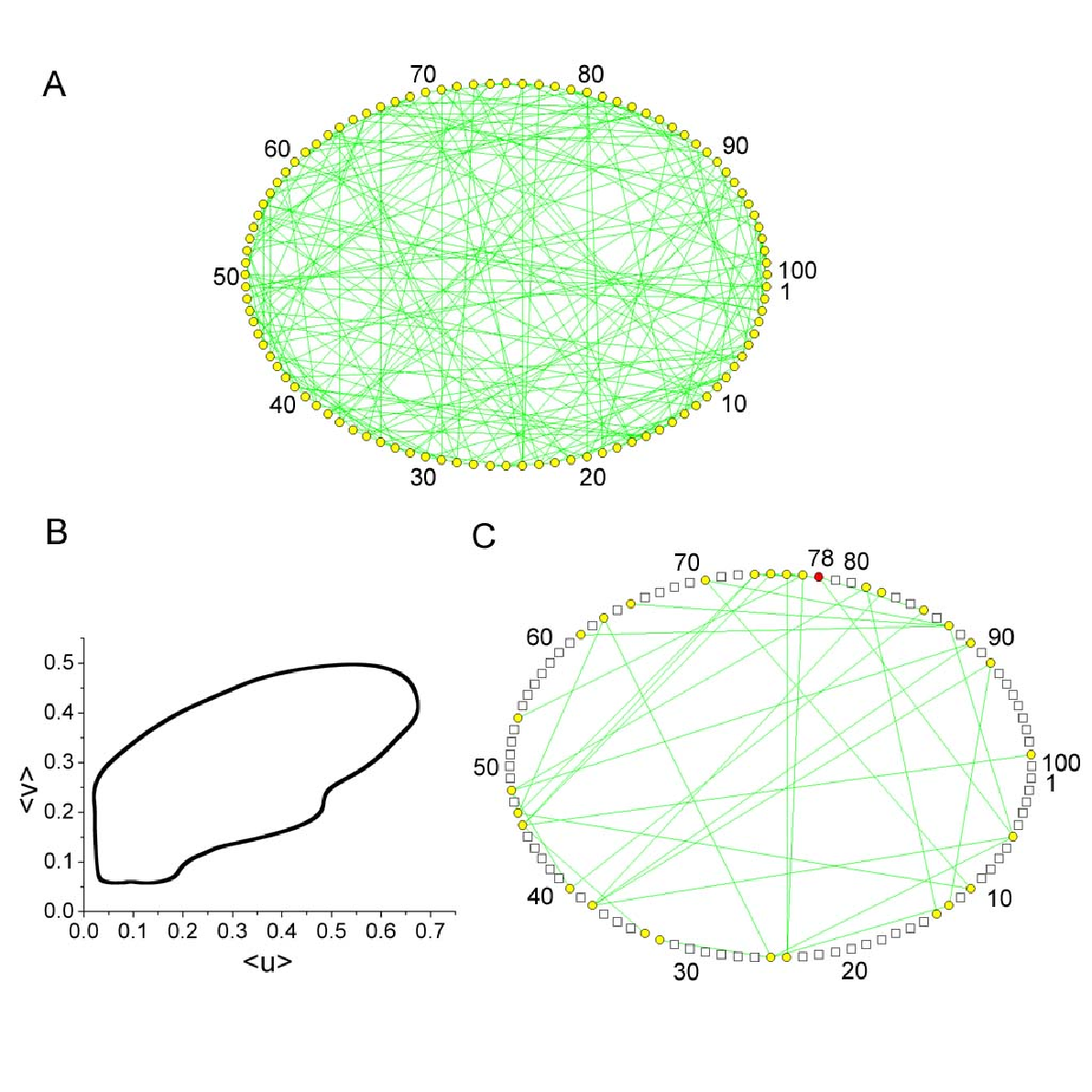}}}\\
\hspace*{\fill}Fig.\ 1

\newpage
{\centering\scalebox{0.8}{\includegraphics{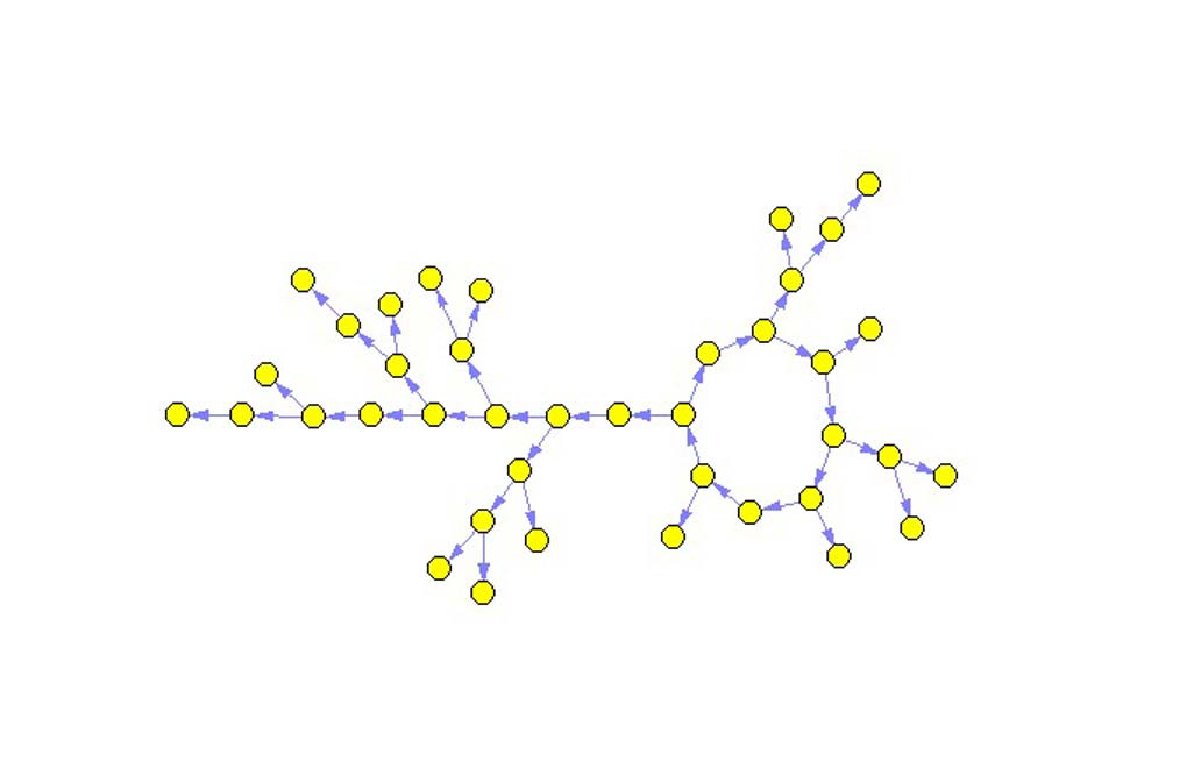}}}\\
\hspace*{\fill}Fig.\ 2

\newpage
{\centering\scalebox{0.8}{\includegraphics{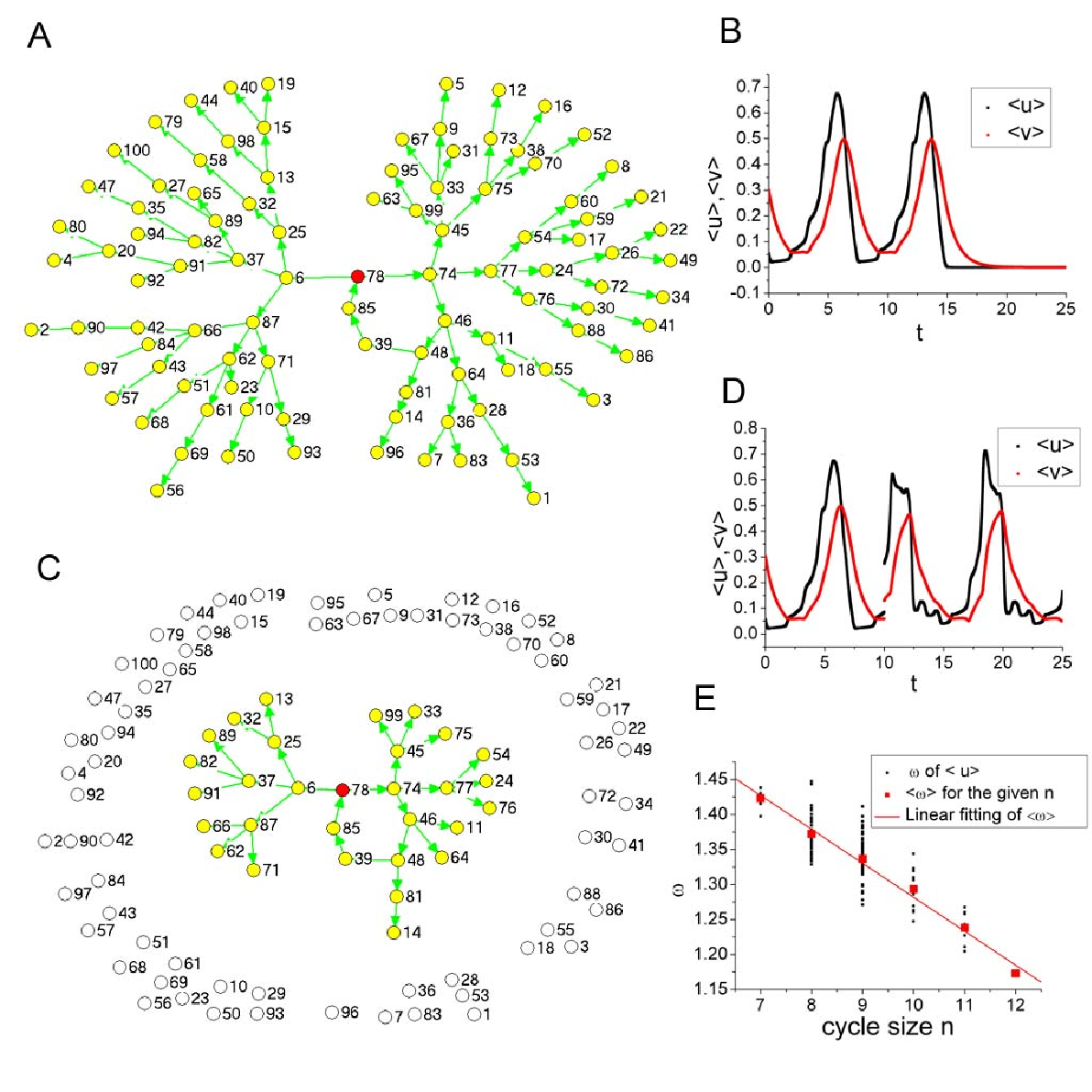}}}\\
\hspace*{\fill}Fig.\ 3

\newpage
{\centering\scalebox{0.8}{\includegraphics{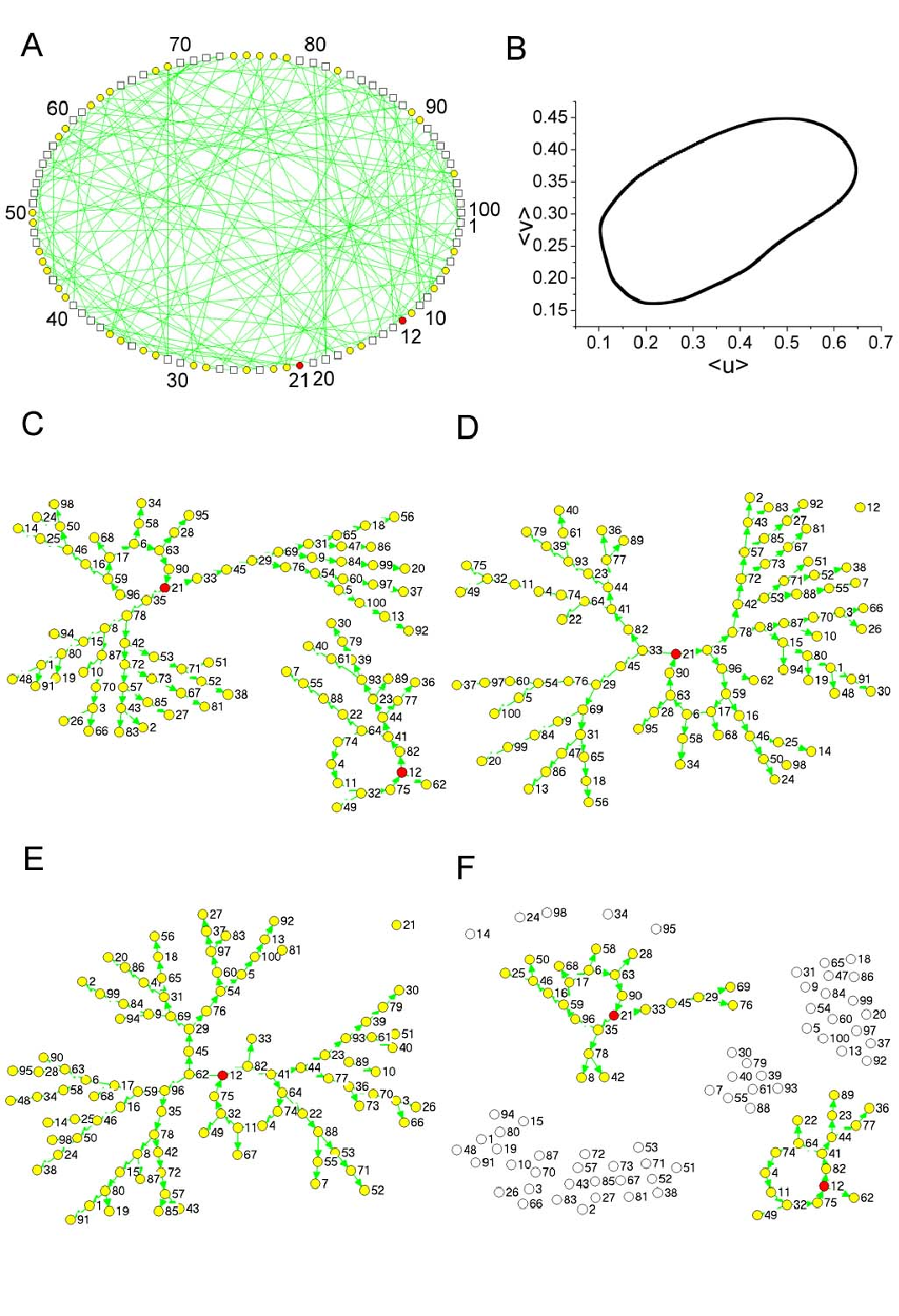}}}\\
\hspace*{\fill}Fig.\ 4

\newpage
{\centering\scalebox{0.8}{\includegraphics{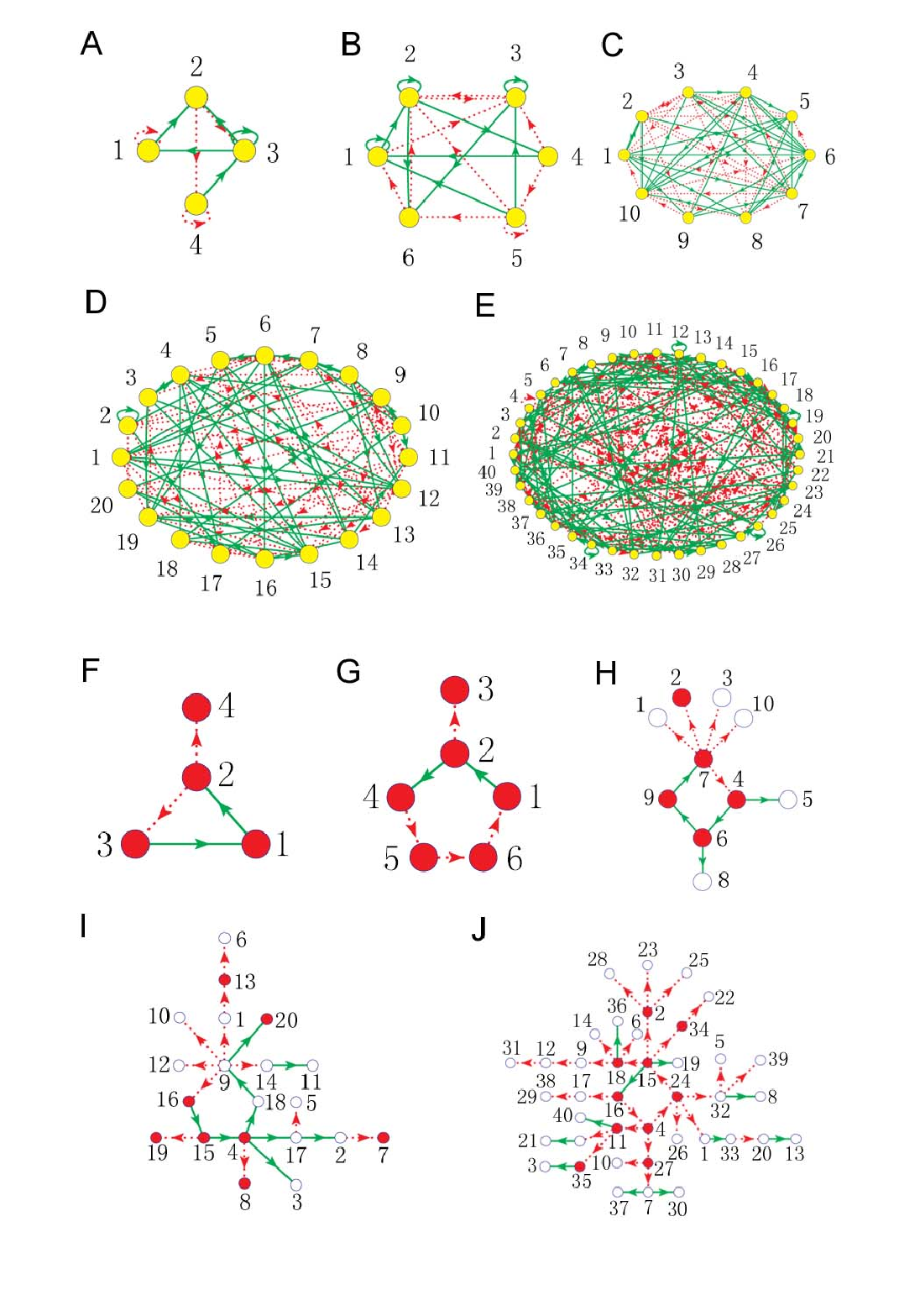}}}\\
\hspace*{\fill}Fig.\ 5

\newpage
{\centering\scalebox{0.8}{\includegraphics{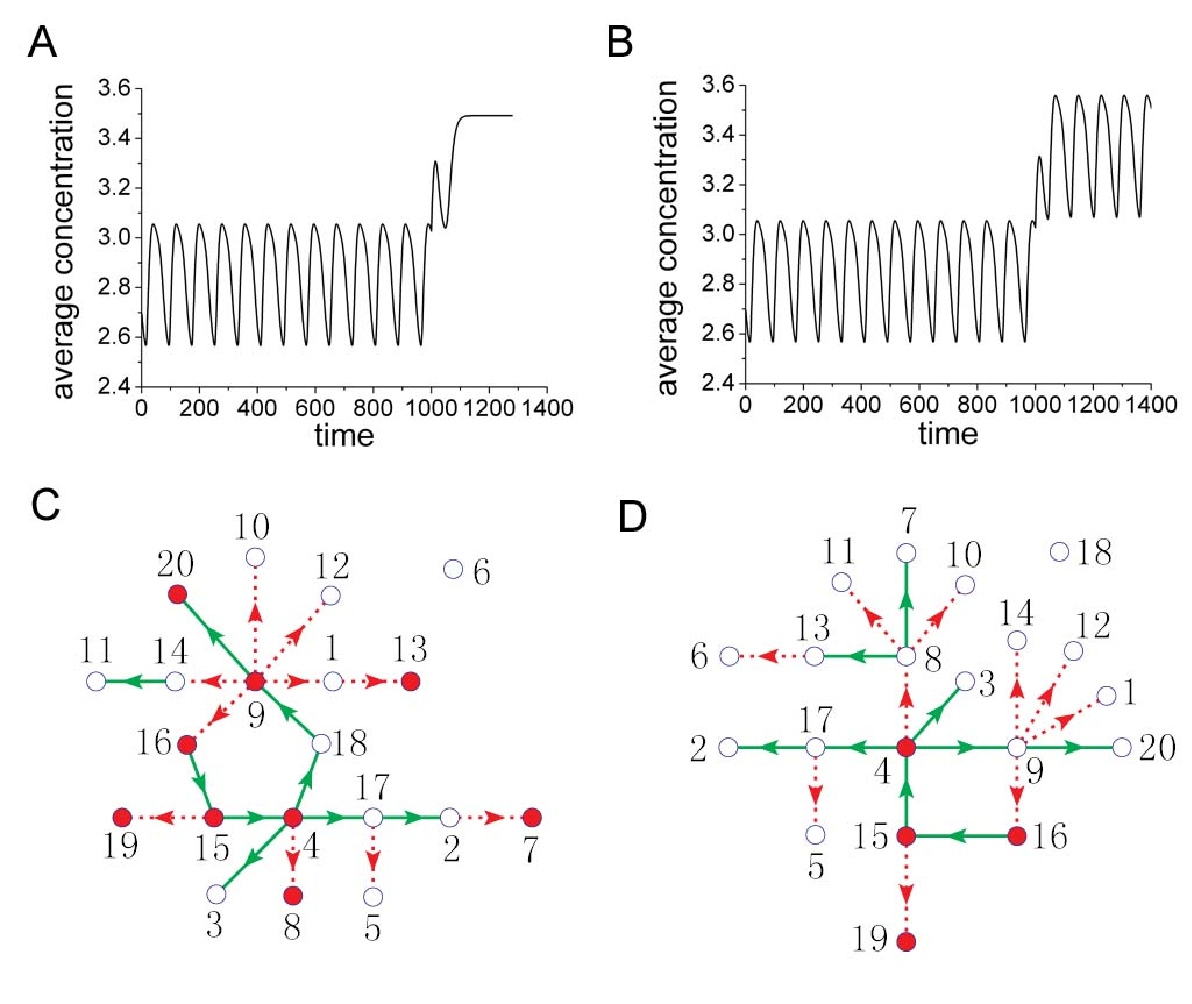}}}\\
\hspace*{\fill}Fig.\ 6

\newpage

{\Large{\bf Supplementary Information}}

\section*{Part 1 Oscillatory excitable
cell networks with random symmetric interactions}

We have investigated a large number of different ECN. First, we have
tested ECNs of Eqs.(1) for different random coupling structures,
different random initial conditions and different system sizes. From
these tests we found many oscillatory configurations. For each
oscillatory network we drew the dominant phase-advanced interaction
pattern, and found that all patterns show BC structures of type
Fig.\ 2.

In Fig.\ S1{\it A} we show an interaction structure of an ECN with
$N=200$, $\nu=3$ and $D_u=1.0$, which is periodically oscillatory
for certain initial conditions. Though the random interaction
network looks even more complicated than Fig.\ 1{\it A}, the reduced
BC patterns are still simple and instructive. In Figs.\ S1{\it B-D}
we show three different BC patterns for three given initial
conditions. These oscillations can be terminated by removing only
one or two key red circle nodes (e.\ g.\ node $30$ for Fig.\ S1B,
nodes $6$ and $38$ for Fig.\ S1{\it C}; and nodes $126$ and $178$
for Fig.\ S1D). A particular case is Fig.\ S1{\it D} where the BC
pattern has a single long $4\pi$ circle consisting of $21$ cells (by
$2k\pi$ circle we mean that phase angle variation in the circle is
$2k\pi$ with wave number $k$). Now we can't suppress oscillations by
removing a single cell. Instead, a pair of cells with phase angle
distance about $2\pi$ should be removed simultaneously. In all the
three cases of Figs.\ S1B-D oscillations can persist and BC circles
cannot be slightly changed after $70\%$ nodes (all at the end of
branches) are removed simultaneously. One of these results is given
in Fig.\ S1{\it E}. It is noticed that in Figs.\ S1{\it B-D} the
nodes immediately upstream to each red node, provide the only
phase-advanced interactions for the corresponding red nodes.
Therefore, removing the nearest upstream node of any red node is
equivalent to remove this downstream red node.

In Fig.S2 we show interesting BC pattern manipulations when
oscillations are not terminated by removing some key nodes. In Fig.\
S2{\it A}(S2{\it B}) we remove a red cell $6$ ($38$) from  Fig.\ S1
{\it C}, and find that the up-left (low-right ) BC circle is
destroyed while the other remaining circle serves as the only source
circle in the new BC pattern. In Fig.\ S2{\it C} (S2{\it D}) we
remove a single red node $126$ (red node $178$) from Fig.\ S1D. Now
we find that the BC circle shrinks from the $4\pi$ circle to a
$2\pi$ loop with the key node $178$ (key node $126$) serving as the
key circle node in the new BC structure. Pictures of Figs.\ 4 and S2
show various structure reorganizations by controlling few key nodes
recognized from the BC patterns. On the other hand, oscillations are
insensitive to the control of various branch nodes.

In Fig.\ S3 we plot various BC patterns for networks with even
larger sizes. In Figs.\ S3{\it A-C} we take $N=400$ for different
interaction structures and different initial conditions. We find BC
patterns with a $2\pi$ circle ({\it A}), $4\pi$ circle ({\it B}),
and two $2\pi$ circles ({\it C}), respectively. The numbers inside
the circles indicate the indexes of the circle nodes while those
outside the circles associated with arrows denote the size of
branches rooted at the given circle nodes. In Figs.\ S3{\it D-F} we
do the same as Figs.\ S3{\it A-C} by taking $N=2500$. Now the
network size is many times larger than Fig.\ 1{\it A}. However, the
prediction of the universal BC structure of Fig.\ 2 is still
verified perfectly. An interesting phenomenon is that we find triple
BC clusters in Fig.\ S3{\it F}.

The BC pattern analysis can be extended to complex networks of
neural cells. Let us consider the network of FHN model(1,2)

$$\frac{du_i}{dt}=\frac{1}{\varepsilon}
(u_i-\frac{u_i^3}{3}-v_i)+D_u \sum_{j=1}^{\nu}(u_{ij}-u_i)\ ,$$
$$\frac{dv_i}{dt}=\varepsilon (u_i+\beta-\gamma v_i),\qquad i=1,2,...,N \quad \eqno (S1)$$
which has been extensively used to describe neural dynamics. For the
given parameter set given here the individual neural cells are
excitable while nonoscillatory. For certain initial preparations the
network of coupled cells can be self-organized to sustained
oscillations. Fig.\ S4{\it A} shows one of such network and Figs.\
S4{\it B} and 4{\it C} present two different periodic orbits of the
same network structure in Fig.\ S4{\it A} for two sets of different
initial conditions. Figs.\ S4{\it D} and S4E present the BC patterns
corresponding to states Figs.\ S4{\it B} and S4{\it C},
respectively. The one circle (Fig.\ S4{\it D}) and two circle (Fig.\
S4{\it E}) BC structures are identified by applying the CR method.

\section*{Part 2 Oscillatory dynamics of genomic regulatory
networks of Eqs.\ (2)}

Firstly, the oscillatory negative feedback loops(NFLs) of GRNs for
$q\leqslant5$ are presented in Fig.\ S5.

In order to define phase-advanced interactions of Eqs.\ (6) and (7),
we should first specify the meaning of phase $\phi_i$ with Fourier
decomposition. With $\phi_i$ known, the phase of interaction from
node $j$ to node $i$ (denoted by $\phi_{j\rightarrow i}$) can be
identified by $\phi_j$ ($\phi_j+\pi$) for positive (negative)
interaction, and the interaction with $0\lesssim \phi_{j\rightarrow
i}-\phi_i\lesssim\frac{\pi}{2}$ is called phase-advanced driving.
For defining the dominant phase-advanced driving of a given node
$i$, we should identify the interaction with the maximum amplitude
among all the phase-advanced interactions of node $i$.

The phase of single node $i$ can be define as follows:

$$sin\phi_i=\frac{\beta_i}{\sqrt{\alpha_i^2+\beta_i^2}},
\qquad \qquad \ \
cos\phi_i=\frac{\alpha_i}{\sqrt{\alpha_i^2+\beta_i^2}},\qquad \quad
\eqno (S2)$$
$$\alpha_i=\frac {2}{T} \int_0^T \Delta x_i sin(\frac{2\pi t }{T})dt,\
\quad \beta_i=\frac {2}{T} \int_0^T \Delta x_i cos(\frac{2\pi
t}{T})dt,$$
$$ \Delta x_i=x_i(t)-\overline{x}_i,\qquad
 \overline{x}_i=\frac{1}{T}\int_0^T x_i(t)dt,\quad i=1,2,\cdots,N. $$

Suppose $A_{j \rightarrow i}(t)$ is the interaction from node $j$ to
node $i$ which can be computed explicitly by linear approximation as

$$A_{j\rightarrow i}(t)=\frac{\partial f_{i}}{\partial x_{j}}
\Delta x_{j} , \eqno (S3)$$
 where $\frac{\partial f_i}{\partial x_j}$
is a constant valued at time averages $\overline{x}_k$
($k=1,2,\cdots,N$). $A_{j \rightarrow i}(t)$ is periodically
oscillatory with period T and zero average. With Eqs.\ (S2) (S3) the
phase $\phi_{j \rightarrow i}$ can be define by $\phi_j$
($\phi_j+\pi$) for positive (negative) interaction, and the
amplitude of $A_{j \rightarrow i}(t)$ is given by

$$ \|A_{j \rightarrow i}\|=\|\frac{\partial f_{i}}{\partial x_{j}}\|\sqrt{\alpha_j ^2 +\beta_j ^2},
\eqno (S4).$$ The dominant phase-advanced driving of node $i$ is the
interaction with the largest amplitude $\|A_{j\rightarrow i}\|$
among all the phase-advanced interactions of node $i$ (i.e, all $j\
$s with $0\lesssim \phi_ {j\rightarrow i}-\phi
_i\lesssim\frac{\pi}{2}$).

The crucial point of the approximations of Eqs.\ (S2)-(S4) is to
neglect the contributions of all high-frequency harmonics of the
interactions and thus the definitions work very well for the cases
of single-peaked periodic motions. Some slight modification is
needed if the motion is multiple-peaked, and these cases are not
considered in the present article. It happens with extremely low
probability that the sum of interactions of a given node obeys the
driving condition Eq.\ (7) while individual interactions do not. In
this case we simply define the interaction with phase nearest to the
region as the dominant interaction. These cases occur always (with
our observations) for the nodes near the branch ends.

\section*{Part 3 Oscillatory dynamics of genomic regulatory networks of type
``OR" interactions}

Eqs.\ (2) consider ``AND " type of joint regulations of activatory
and repressive regulators, represented by the productive formula. In
realistic regulatory networks there also exist joint regulations of
type ``OR ". Typical mathematical formulas of a network of this type
(3,4) with size $N$ are as follows:

$$ \frac{dx_i}{dt}=\gamma_i (1-x_i)+\sum\limits_{j=1}^N f_{ij} \ ,
\qquad i=1,2,\cdots,N \ ,\eqno (S5) $$

\[
 f_{ij} = \left\{
  \begin{array}{ll}
  0 &\mbox{no coupling}\ , \\
  (1-x_i)\frac{k_{ij}\, x_j^{h_i}}{K_i^{h_i}+\,{x_j}^{h_i}} &\mbox{positive coupling}\ , \\
  -x_i\frac{k_{ij}\, x_{j}^{h_i}}{K_i^{h_i}+\,{x_j}^{h_i}} &\mbox{negative coupling}\ .
  \end {array}
  \right.
\]
Both Eqs.\ (2) and (S5) are approximations of more realistic as well
as more complicated regulatory networks mixing ``AND" and ``OR"
dynamics.

Since in Eqs.\ (S5) each node does not oscillate individually and
positive (negative) regulations tend to increase (reduce)
monotonously the concentration of the protein interacted, the phase
relations analyzed in the article, are satisfied entirely. In
particular, the phase and amplitude of couplings are given by Eqs.\
(6) and (S4), respectively, with $f_{i}$ replaced by $\sum_{j=1}^N
f_{ij}$. For mathematical simplicity we again take identical
parameters in Eqs.\ (S5) for all nodes $h_i=h$, $k_{ij}=k$, $K_i=K$,
$i=1,2,\ldots,N$, except $\gamma_{i=1,\ldots,N-1}=\gamma$,
$\gamma_N=1.0$.

In Figs.\ S6 we do the same as Fig.\ 5 with Eqs.\ (2) replaced by
the ``OR" dynamics of Eqs.\ (S5). It is really striking that though
the coupling dynamics of Eqs.\ (S5) is different from that of Eqs.\
(2) considerably, BC patterns of Eqs.\ (S5) have essentially the
same structures as Fig.\ 5, and all the features predicted in Fig.\
2 are confirmed perfectly in Fig.\ S6. Moreover, the prediction of
oscillation sources (one of the oscillatory NFLs in Fig.\ S5) is
also fully verified by all the BC circles obtained from Eqs.\ (S5).
We have also made statistics of the circuits of Eqs.\ (S5) with huge
number of tests by varying coupling structures, parameter values and
initial conditions ($10^6$ for 3-node, $10^4$ for 4, 5-node, $10^3$
for 10, 20-node). We find that the predictions on BC structures in
Fig.\ 2 and Fig.\ S5 are well confirmed by all tests of oscillatory
circuits with no exception.

\noindent {\bf Fig.\ S1.} Oscillatory behaviors of a complex ECN of
Eqs.\ (1) with $N=200$, $\nu=3$, $D_u=1.0$. All other parameters are
the same as Fig.\ 1(A). {\it(A)} Interaction structure under
consideration. {\it(B)}-{\it(D)} BC patterns of periodically
oscillatory states of ECN (A) for three different initial
conditions. All red nodes have the same meanings as those in
article. {\it(B)} BC pattern with a $2\pi$ circle. {\it(C)} BC
pattern with two source circles. In order to suppress the
oscillation one has to remove the two red nodes $6$ and $38$
simultaneously, destroying both source circles. {\it(D)} BC pattern
with with a $4\pi$ (wave number $k=2$) circle of circle size $n=21$.
We find two red nodes in the circle separated by about $2\pi$
distance. In order to suppress oscillation we have to remove both
red nodes $126$ and $178$ simultaneously. {\it(E)} The BC pattern
constructed for the new oscillatory state after removing $140$ nodes
from (D). Oscillation persists and the original BC circle remains
unchanged after $70\%$ nodes are removed. Similar results (not
shown) can be obtained also for BC patterns of (B) and (C).

\noindent {\bf Fig.\ S2.} BC pattern manipulations. {\it(A)(B)} BC
patterns after removing a red node $6$ and the other red node $38$
from Fig.\ S1(C), respectively. In both cases, the circle containing
the removed node is destroyed and the remained circle serves as the
unique oscillation source of the network. {\it(C)(D)} BC patterns
with red nodes $126$ and $178$ removed from Fig.\ S1(D),
respectively. Now the $4\pi$ BC circle of Fig.\ S1(D) shrinks to a
$2\pi$ circle with the remaining red node being the key circle node
in the new BC pattern.

\noindent  {\bf Fig.\ S3.} BC patterns of Eqs.\ (1) for different
system sizes with $\nu =3$, $D_u=1.0$. All other parameters are the
same as Fig.\ 1(A). Interaction structures and initial conditions
are chosen randomly. {\it(A)}-{\it(C)} BC patterns with $N=400$ for
different states. Numbers inside the circle indicate the
corresponding indexes of corresponding nodes, and the numbers
outside the circle, associated to circle nodes by arrows, denote the
sizes of the branches rooted at the given nodes. All nodes without
arrows have no downstream branch. Single $2\pi$ BC circle; single
$4\pi$ circle and double-circle structure are observed in (A)(B)(C),
respectively. {\it(D)}-{\it(F)} The same as (A)-(C) with $N=2500$.
Single-, double- and triple-circle BC structures are identified in
(D)-(F), respectively.

\noindent {\bf Fig. S4.} Oscillatory complex network of FHN cells
(Eqs.\ (S1)) and some examples of BC patterns. The parameters are
set as $\varepsilon=0.2$, $\gamma=0.5$, $\beta=1.0$. All couplings
with coupling strength $D_u=0.1$ are randomly chosen between $N$
nodes each having coupling degree $\nu$. {\it(A)} An example of FHN
network with $N=100$, $\nu =3$. {\it(B)(C)} Two periodic orbits of
network (A) for two different initial conditions. {\it(D)(E)} The BC
patterns corresponding to the states (B) and (C), respectively. The
red nodes have the same meanings as Figs.\ 3(A) and 4(C).

\noindent {\bf Fig. S5.} All oscillatory $q$-node NFLs of Eqs.\ (2)
for $q\leq 5$. The solid green (red dashed) line denotes positive
(negative) regulation. {\it(A)} $2$-node NFL. {\it(B)} $3$-node NFL
with three negative interactions. {\it(C)} $3$-node NFL with a
single negative interaction. {\it(D)} $4$-node NFL with three
negative interactions. {\it(E)} $4$-node NFL with a single negative
interaction. {\it(F)} $5$-node NFL with a single negative
interaction. {\it(G)} $5$-node NFL with three negative interactions
(Type I). {\it(H)} $5$-node NFL with three negative interactions
(Type II). {\it(I)} $5$-node NFL with five negative interactions.
All these oscillatory NFLs have been observed in various $BC$
patterns.

\noindent {\bf Fig. S6.} Numerical results on oscillatory GRNs of
Eqs.\ (S5) (type ``OR").  Homogeneous parameter sets are used with
$h_i=h$, $k_{ij}=k$, $K_i=K$, $i=1,2,\ldots,N$, except
$\gamma_{i=1,\ldots,N-1}=\gamma$, $\gamma_N=1.0$. We use two sets of
parameters in simulation. In Figs.\ (F), (G), (J) $h=3$,
$k=734.5525513$, $K=3.5531352$, $\gamma=0.8991836$; in Figs.\ (H),
(I) $h=4$, $k=753.6924438$, $K=0.6794546$, $\gamma=0.1966502$. These
parameters are randomly taken in the available ranges specified in
Science report (3). In all systems unique periodically oscillatory
solutions are asymptotically approached from arbitrary initial
conditions. {\it(A)}-{\it(E)} Interaction structures of GRNs of
different size $N$ and different number $I$ of interactions.
{\it(A)} $N=4$, $I=9$. {\it(B)} $N=6$, $I=28$. {\it(C)} $N=10$,
$I=56$. {\it(D)} $N=20$, $I=109$. {\it(E)} $N=40$, $I=222$.
{\it(F)}-{\it(J)} BC patterns corresponding to (A)-(E),
respectively. Red and empty nodes have the same meanings as in
Figs.\ 5 (F)-(J).

\newpage
{\centering\scalebox{0.8}{\includegraphics{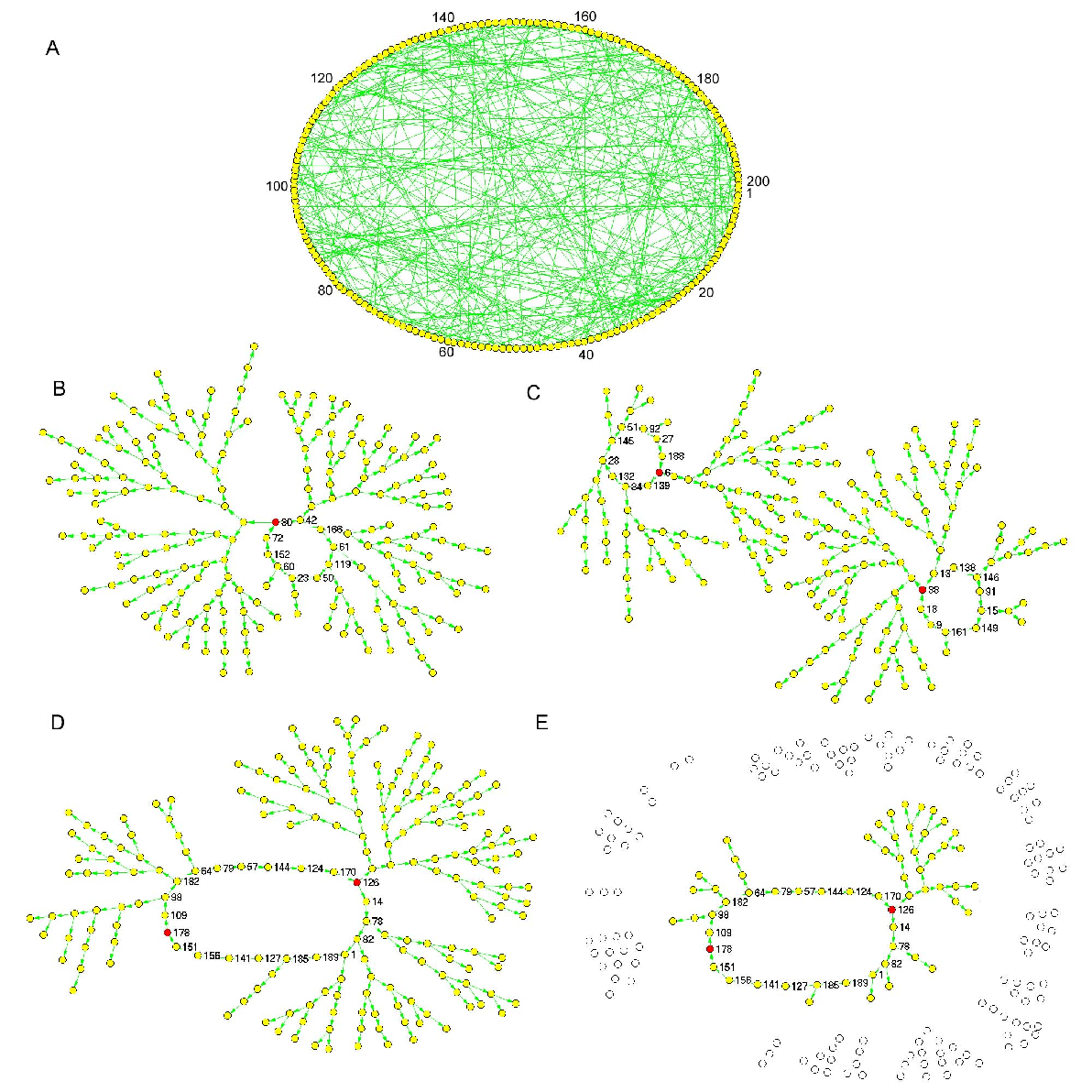}}}\\
\hspace*{\fill}Fig.\ S1

\newpage
{\centering\scalebox{0.8}{\includegraphics{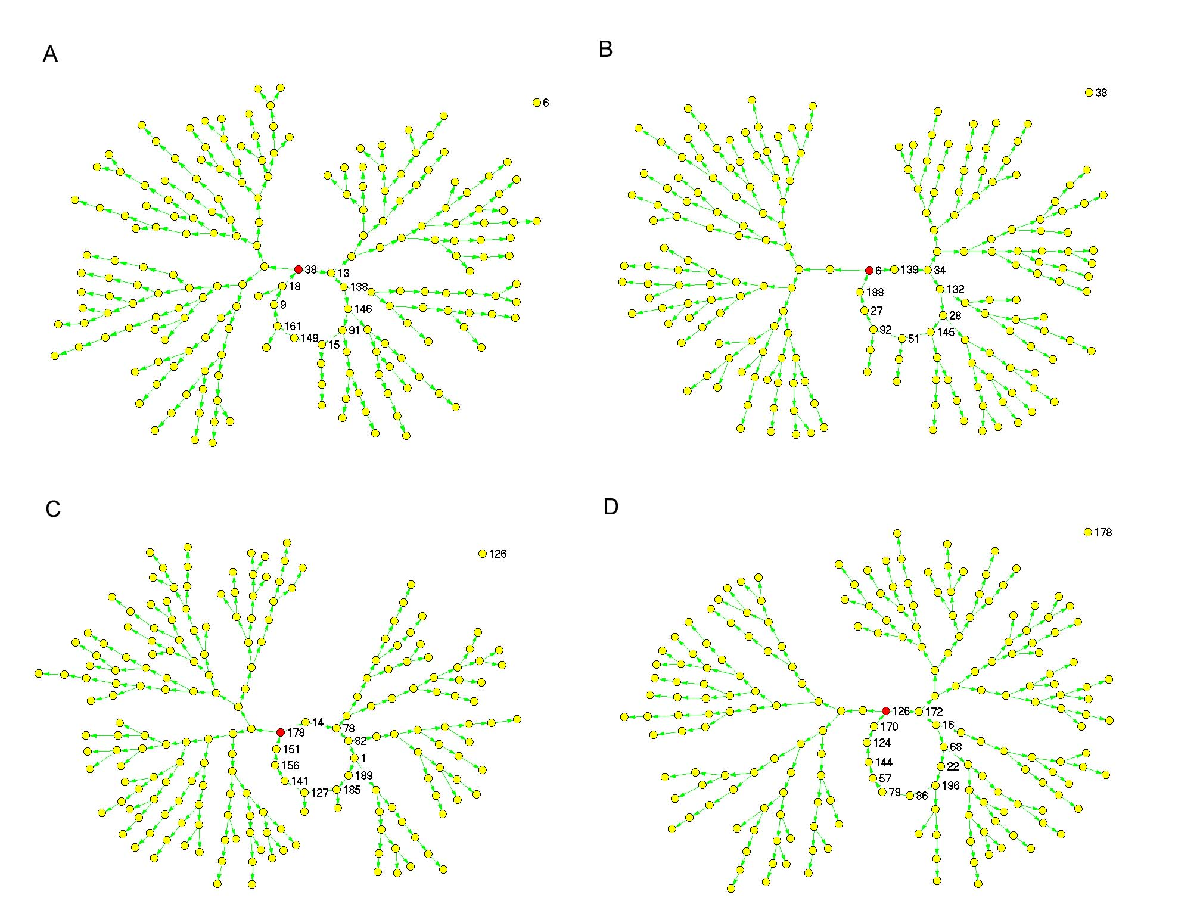}}}\\
\hspace*{\fill}Fig.\ S2

\newpage
{\centering\scalebox{0.8}{\includegraphics{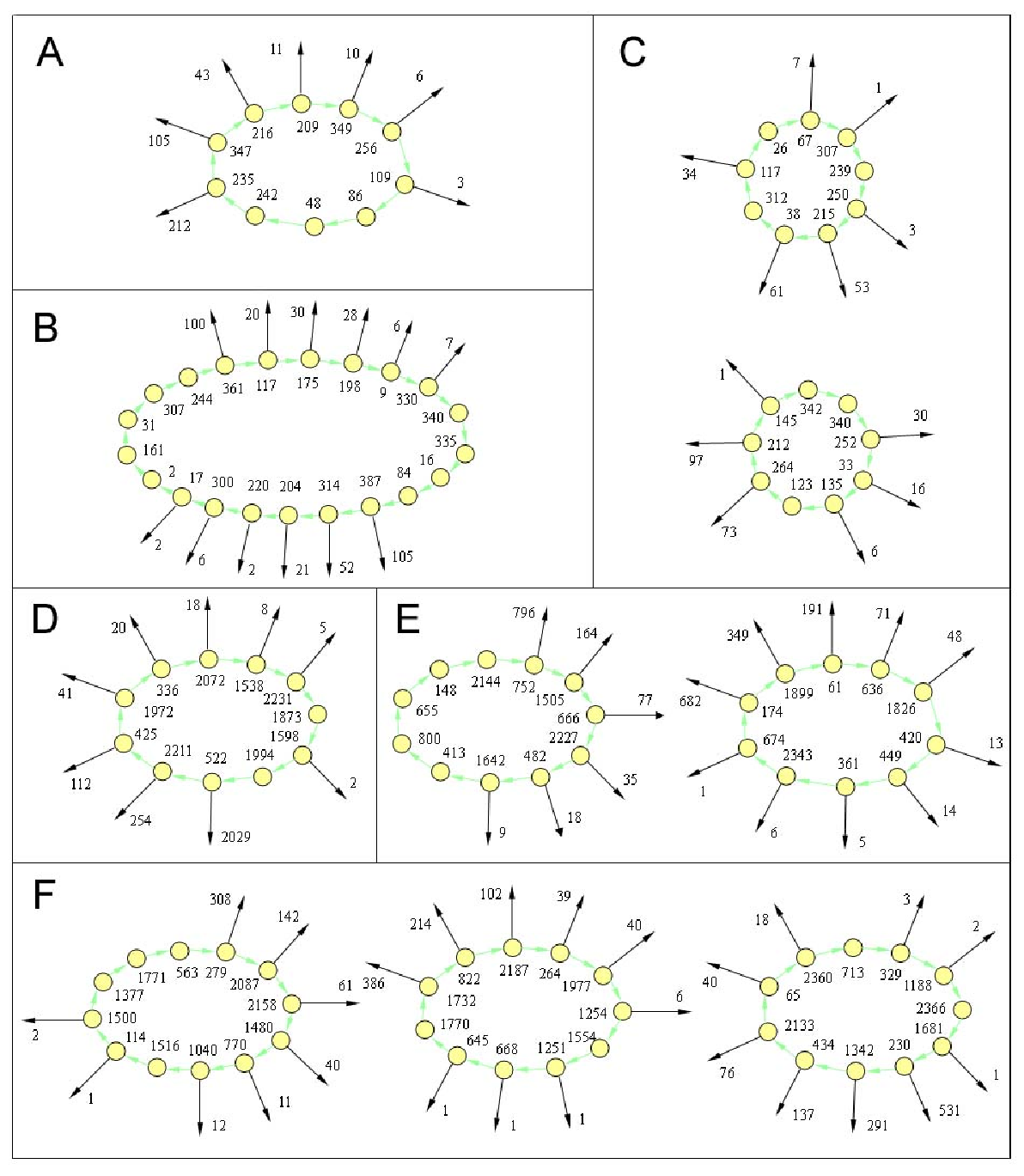}}}\\
\hspace*{\fill}Fig.\ S3

\newpage
{\centering\scalebox{0.8}{\includegraphics{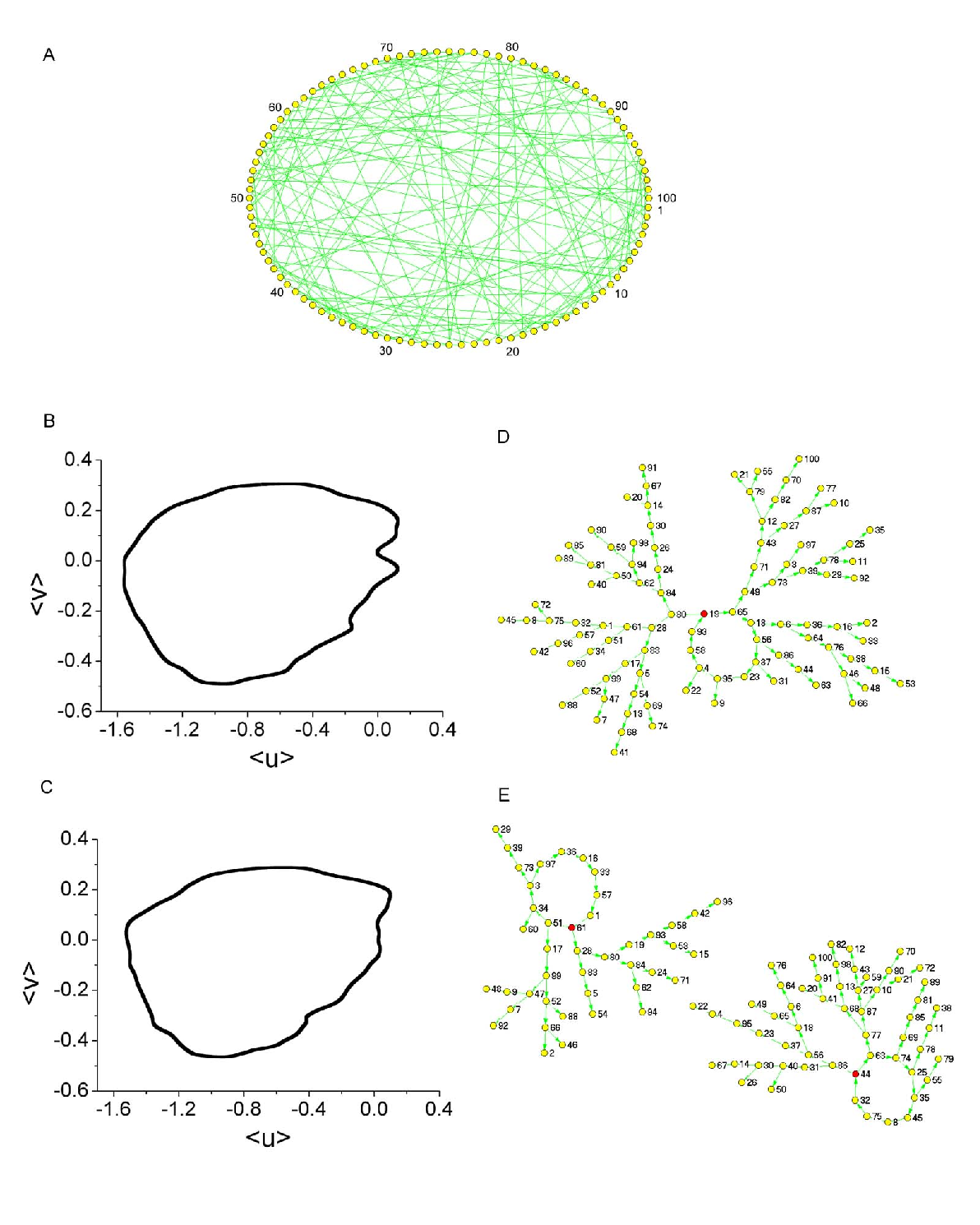}}}\\
\hspace*{\fill}Fig.\ S4

\newpage
{\centering\scalebox{0.8}{\includegraphics{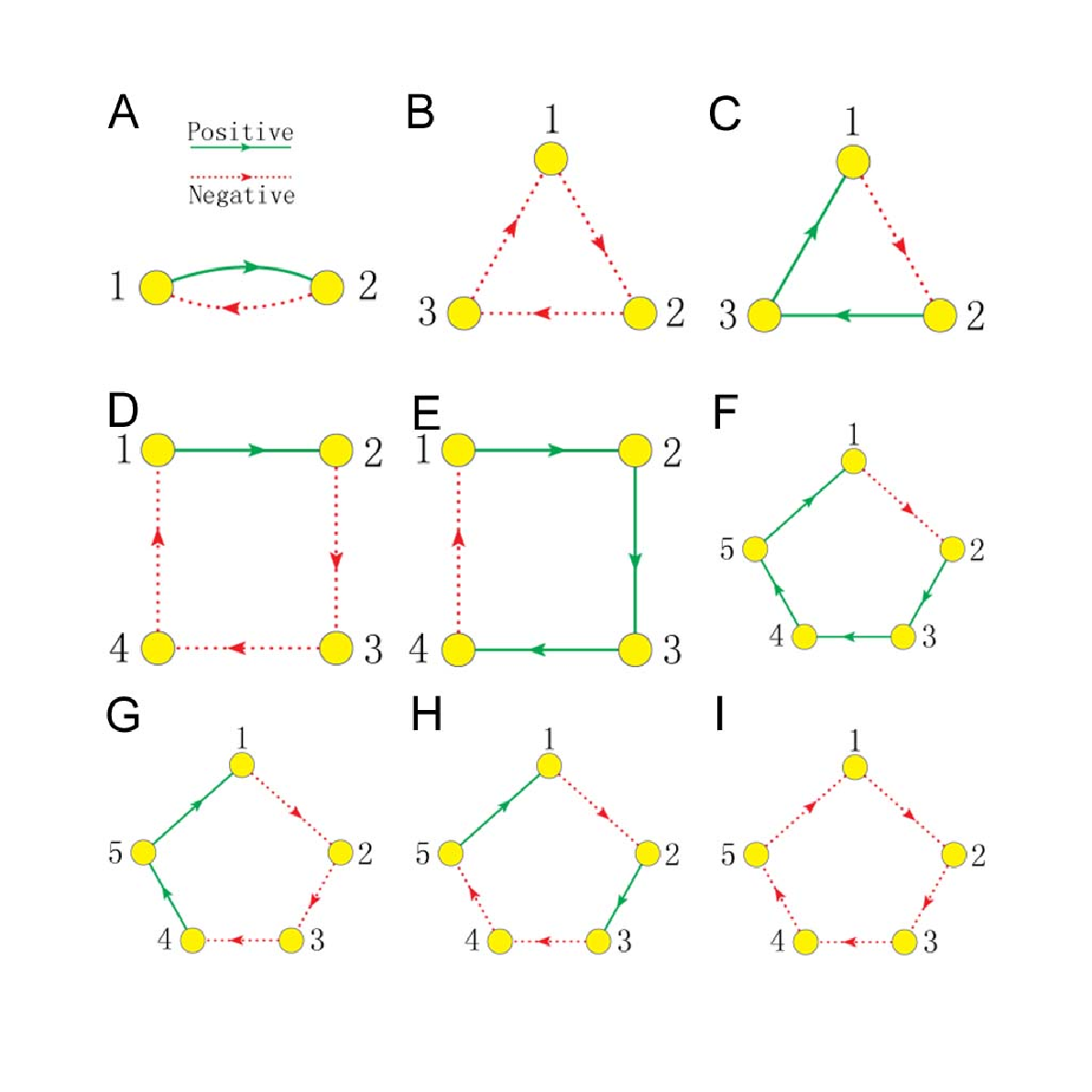}}}\\
\hspace*{\fill}Fig.\ S5

\newpage
{\centering\scalebox{0.8}{\includegraphics{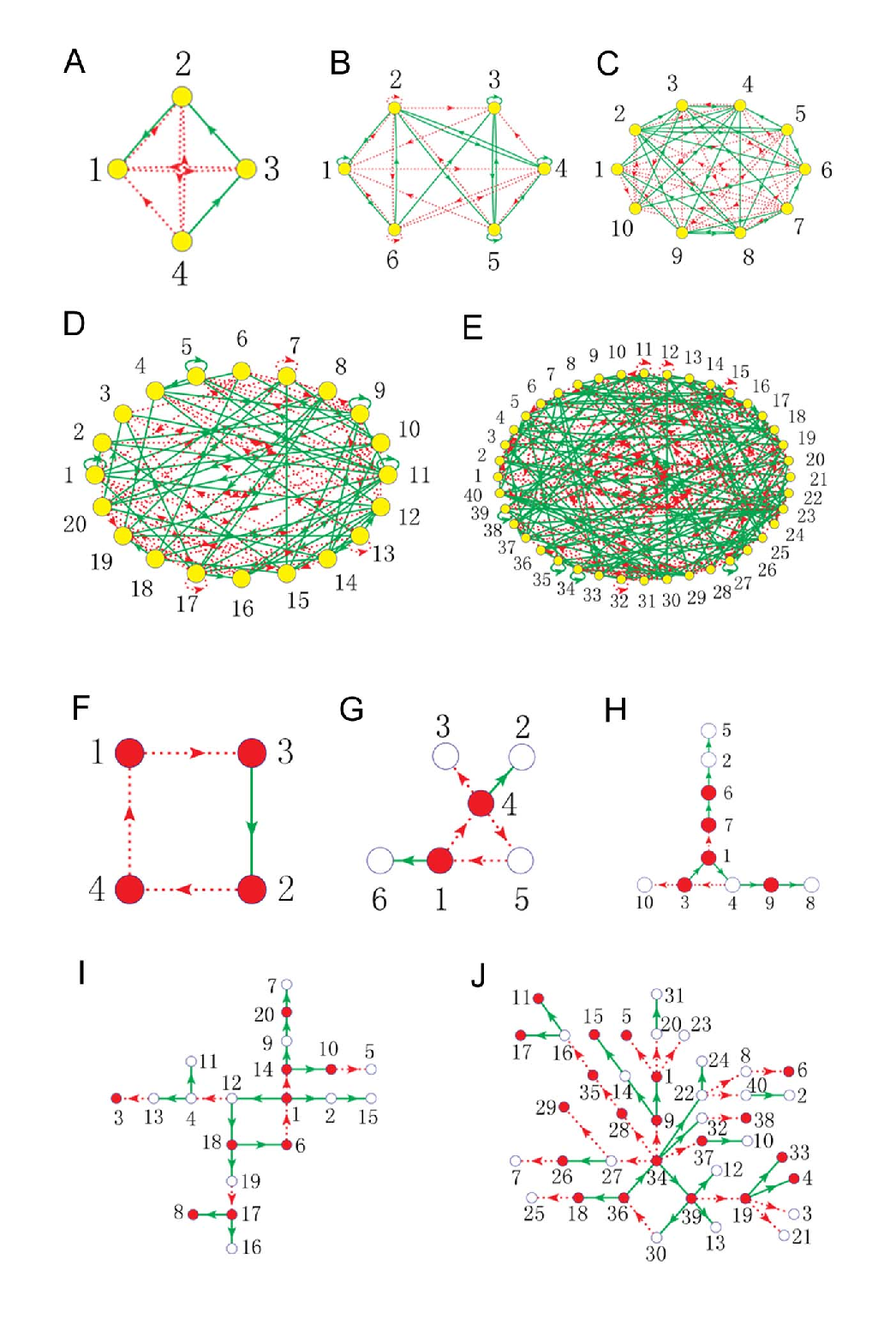}}}\\
\hspace*{\fill}Fig.\ S6

\clearpage


\begin{thebibliography}{2}
\bibitem{osc_1} Buzs$\acute{a}$ki G, Draguhn A (2004)
  Neuronal oscillations in cortical networks.
  {\it Science\/} 304:1926-1929.
  \bibitem{osc_2} Sanchez-Vives M V, McCormick D A (2000)
  Cellular and network mechanisms of rhythmic recurrent activity in neocortex.
  {\it Nature Neurosci.\/} 3:1027-1034.
  \bibitem{osc_3} Steriade M, Timofeev I (2003)
  Neuronal plasticity in thalamocortical networks during sleep and waking oscillations.
  {\it Neuron\/} 37:563-576.
  \bibitem{osc_4} Selverston A I, Moulins M (1985)
  Oscillatory neural networks.
  {\it Ann. Rev. Physiol.\/} 47:29-48.
  \bibitem{num_1} Lewis T J, Rinzel J (2000)
  Self-organized synchronous oscillations in a network of
  excitable cells coupled by gap junctions.
  {\it Network: Comput.\ Neural Syst.\/} 11:299-320.
  \bibitem{num_2} Sinha S, Saram$\ddot{a}$ki J, Kaski K (2007)
  Emergence of self-sustained patterns in small-world excitable
  media. {\it Phys.\ Rev.\ E\/} 76:R015101.
  \bibitem{num_3} Roxin A, Riecke H, Solla S A (2004)
  Self-sustained activity in a small-world network
  of excitable neurons. {\it Phys.\ Rev.\ Lett.\/} 92:198101.
  \bibitem{sino_1} Irisawa H, Brown H F, Giles W (1993)
  Cardiac pacemaking in the sinoatrial node.
  {\it Physiol.\ Rev.\/} 73:197-227.
  \bibitem{sino_2} Boyett M R, Honjo H, Kodama I (2000)
  The sinoatrial node, a heterogeneous pacemaker structure.
  {\it Cardiovasc. Res.\/} 47:658-687.
  \bibitem{sino_3} Mandel W, Hayakawa H, Danzig R,
  Marcus H S (1971) Evaluation of sino-atrial node function in man by
  overdrive suppression. {\it Circulation\/} 44:59-66.

  \bibitem{grn_1} Tsai T Y C, {\it et al.\/} (2008) Robust,
  tunable biological oscillations from interlinked positive
  and negative feedback loops. {\it Science\/} 321:126-129.
  \bibitem{grn_2} Pomerening J R, Kim S Y, Ferrell J E, Jr (2005)
  Systems-level dissection of  the cell-cycle oscillator:
  bypassing positive feedback produces damped oscillations.
  {\it Cell\/} 122:565-578.
  \bibitem{grn_3} Elowitz M B, Leibler S (2000) A synthetic
  oscillatory network of transcriptional regulators.
  {\it Nature\/} 403:335-338.
  \bibitem{grn_4} Glossop N R J, Lyons L C, Hardin P E (1999)
  Interlocked feedback loops within the drosophila circadian
  oscillator. {\it Science\/} 286:766-768.
  \bibitem{grn_5} Hirata H,{\it et al.} (2002) Oscillatory
  expression of the bHLH Factor Hes1 regulated by a negative
  feedback loop. {\it Science \/} 298:840-843.
  \bibitem{grn_6} Geva-Zatorsky N, {\it et al.\/} (2006)
  Oscillations and variability in the p53 system.
  {\it Mol.\ Syst.\ Biol.\/} 13:msb4100068E1.
  \bibitem{grn_7} Rust M J, Markson J S, Lane W S,
  Fisher D S, O'Shea E K (2007) Ordered phosphorylation
  governs oscillation of a three-protein circadian clock.
   {\it Science\/} 318:809-812.

  \bibitem{Bar_1} B$\ddot{a}$r M, Eiswirth M (1993) Turbulence
  due to spiral breakup in a continuous excitable medium.
  {\it Phys.\ Rev.\ E\/} 48:R1635.
  \bibitem{num_4} Jahnke W, Winfree A T (1991)
  A survey of spiral-wave behaviors in the Oregonator
  model. {\it Int.\ J.\ Bifur.\ Chaos\/} 1:445--466.
  \bibitem{neu_1} Fitzhugh R (1961) Impulses and physiological
  states in theoretical models of nerve membrane.
  {\it Biophys.\ J.\/} 1:445-466.
  \bibitem{mod_1} Ishihara S, Fujimoto K, Shibata T (2005)
  Cross talking of network motifs in gene regulation
  that generates temporal pulses and spatial stripes.
  {\it Genes to Cells\/} 10:1025-1038.
  \bibitem{mod_2} Pigolotti S, Krishna S, Jensen M H (2007)
  Oscillation patterns in negative feedback loops.
  {\it PNAS\/} 104:6533-6537.
  \bibitem{mod_3} Li C, Chen L, Aihara K (2006)
  Stability of genetic networks with sum regulatory logic:
  Lur'e system and LMI approach.
  {\it IEEE Trans.\ Cir.\ Sys.\/} 53:2451-2458.
  \bibitem{ODE}  Mangan S, Zaslaver A, Alon U (2003)
  The coherent feedforward loop serves as a sign-sensitive
  delay element in transcription networks.
  {\it J.\ Mol.\ Biol.\/} 334:197-204.

  \bibitem{neu_2} Hodgkin A L, Huxley A F (1952)
  A quantitative description of membrane current and
  its application to conduction and excitation in nerve.
  {\it J.\ Physiol.\/} 117:500-544.
  \bibitem{neu_3} Izhikevich E M (2000)
  Neural excitability, spiking and bursting.
  {\it Int.\ J.\ Bifur.\ Chaos\/} 10:1171-1266.
  \bibitem{neu_4} Izhikevich E M (2004)
  Which model to use for cortical spiking neurons.
  {\it IEEE Trans.\ on Neural Networks\/} 15:1063-1070.
  \bibitem{motif} Alon U (2007)
  Network motifs: theory and experimental approaches.
  {\it Nature Rev.\ Genet.\/} 8:450-461.
  \bibitem{Or} Buchler N E, Gerland U, Hwa T (2003)
  On schemes of combinatorial transcription logic.
  {\it PNAS\/} 100:5136-5141.
  \bibitem{CPG_1} Grillner S, Wallen P (1985)
  Central pattern generators for locomotion,
  with special reference to vertebrates.
  {\it Ann.\ Rev.\ Neurosci.\/} 8:233-261.
  \bibitem{CPG_2} Marder E, Bucher D (2001)
  Central pattern generators and the control of
  rhythmic movements. {\it Curr.\ Biol.\/} 11:R986-R996.
  \bibitem{CPG_3} Kiehn O, Butt S J B (2003)
  Physiological, anatomical and genetic identification of
  CPG neurons in the developing mammalian spinal cord.
  {\it Prog.\ Neurobiol.\/} 70:347-361.
  \bibitem{CPG_4} Yuste R, MacLean J N, Smith J, Lansner A (2005)
  The cortex as a central pattern generator.
  {\it Nature Rev.\ Neurosci.\/} 6:477-483.

\end{thebibliography}

\begin{thebibliography}{1}
  \bibitem{S1} Fitzhugh R (1961) Impulses and physiological
  states in theoretical models of nerve membrane.
  {\it Biophys.\ J.\/} 1:445-466.
  \bibitem{S2} Izhikevich E M (2004)
  Which model to use for cortical spiking neurons.
  {\it IEEE Trans.\ on Neural Networks\/} 15:1063-1070.
  \bibitem{S3} Elowitz M B, Leibler S (2000) A synthetic
  oscillatory network of transcriptional regulators.
  {\it Nature\/} 403:335-338.
  \bibitem{S4}  Buchler N E, Gerland U, Hwa T (2003)
  On schemes of combinatorial transcription logic.
  {\it PNAS\/} 100:5136-5141.
\end{thebibliography}
\end{document}